\documentclass[aps,prl,twocolumn,showpacs]{revtex4}
\usepackage{graphics,epsfig,graphicx}
\usepackage{color}
\usepackage{makeidx}
\usepackage{latexsym}
\usepackage{calrsfs}
\usepackage{hyperref}

\begin{document}

\title{Resonant interaction of a single atom with single photons \\
from a down-conversion source}

\date{\today}

\author{C. Schuck\footnote{These authors contributed
equally to the work.}, F. Rohde$^*$, N. Piro$^*$, M.
Almendros$^*$, J.~Huwer, M.~W.~Mitchell,
M.~Hennrich\footnote{Current address: Institut f\"{u}r
Experimentalphysik, Universit\"{a}t Innsbruck, Austria}, A.
Haase\footnote{Current address: Dipartamento di Fisica,
Universit\'{a} di Trento, Italy}, F. Dubin, J. Eschner}

\affiliation{ICFO - Institut de Ciencies Fotoniques, Mediterranean Technology Park, 08860
Castelldefels (Barcelona), Spain
}


\begin{abstract}
We observe the interaction of a single trapped calcium ion with single
photons produced by a narrow-band, resonant down-conversion source [A.
Haase et al., Opt. Lett. \textbf{34}, 55 (2009)], employing a quantum jump
scheme. Using the temperature dependence of the down-conversion spectrum
and the tunability of the narrow source, absorption of the down-conversion
photons is quantitatively characterized.
\end{abstract}

\maketitle

At the level of single particles, the quantum nature of light-matter
interaction becomes manifest, and the absorption and emission of single
photons by single atoms is one of the key physical processes on which
quantum optics is built. Seminal examples of phenomena in this respect are
photon anti-bunching \cite{Kimble1977PRLv39p691, Diedrich1987PRLv58p203},
quantum jumps \cite{Nagourney1986PRLv56p2797, Sauter1986PRLv57p1696,
Bergquist1986PRLv57p1699, Gleyzes2007Nv446p297}, Jaynes-Cummings dynamics
\cite{Brune1996PRLv76p1800}, and atom-photon entanglement
\cite{Blinov2004Nv428p153, Volz2006PRLv96p30404, Wilk2007Sv317p488}.

At the same time, important applications of quantum optics, in particular
in quantum optical information technology and in quantum metrology, are
based on atom-photon interaction at the single particle level. The most
precise clock is realized with a single laser-excited trapped ion
\cite{Rosenband2008Sv319p1808}, and strings of trapped ions have been
shown to be promising systems for implementing quantum logical algorithms
\cite{Schmidt-Kaler2003Nv422p408, Leibfried2003Nv422p412,
Haeffner2005Nv438p643, Leibfried2005Nv438p639, Benhelm2008NPv4p463} as
well as quantum networks \cite{Moehring2007Nv449p68}.

A key step in converting quantum optical phenomena into quantum technology tools is the
control of the processes at all levels, i.e.\ of the atomic internal (electronic) and
external (motional) state, and of the parameters of the photons, including their spatial
and temporal shape, their polarization, and, ideally, their arrival times. Two major
strategies may be distinguished how such control is obtained: on the one hand through
deterministic operations, whereby typically photons are confined and controlled by
high-finesse cavities, on the other hand through probabilistic operations, whereby some
experimental signal indicates that the desired interaction process has occurred. Further
approaches include collective effects in atomic samples \cite{Thompson2006Sv313p74,
Chaneli`ere2005Nv438p833, Chou2007Sv316p1316}, optics with very high opening angles
\cite{Sondermann2007APBv89p489}, or temporal and geometrical pulse shaping; the state-of
the-art of the latter is summarized in Ref.~\cite{Stobi´nska2009Ev86p14007}.

With deterministic atom-light interaction in cavities, important results have been
achieved such as single- \cite{McKeever2004Sv303p1992, Kuhn2002PRLv89p67901,
Keller2004Nv431p1075, Barros2009avp} and entangled-photon \cite{Weber2009PRLv102p30501}
creation and photon turnstile operation \cite{Dayan2008Sv319p1062}. A prominent example
for probabilistic operations is the experimental creation of remote atom-atom
entanglement \cite{Moehring2007Nv449p68}, and many more ideas exist including quantum
repeaters \cite{Briegel1998PRLv81p5932} and all-optical quantum computing
\cite{Knill2001Nv409p46}. In terms of control of atomic states, single trapped ions have
produced very advanced results such as quantum coherence on time scales of seconds
\cite{Roos2004PRLv92p220402, Langer2005PRLv95p60502}, two- and three-ion quantum gates
\cite{Schmidt-Kaler2003Nv422p408, Leibfried2003Nv422p412, Monz2009PRLv102p40501,
Chiaverini2005Sv308p997}, and multi-ion entanglement \cite{Haeffner2005Nv438p643,
Leibfried2005Nv438p639}. In terms of control of individual photons, spontaneous
parametric down-conversion (SPDC) sources produce entangled photon pairs at high
fidelities and rates \cite{Fedrizzi2007OEv15p15377, Wolfgramm2008OEv16p18145}, and serve
as "heralded" single-photon sources \cite{Hong1986PRLv56p58}.

In this context we report the observation of interaction between a single
atom and single photons from a spontaneous parametric down-conversion
source. The atom in our experiment is a single $^{40}$Ca$^+$ ion, trapped
in a linear Paul trap and cooled by continuous laser excitation; the
photons are produced by a SPDC source, described in more detail in
Refs.~\cite{Haase2009OLv34p55, Piro2009JPBAMOPv42p114002}, which is tuned
to provide entangled photon pairs in the wavelength range of the
$4{\textrm D}_{3/2} - 4{\textrm P}_{3/2}$ transition in Ca$^+$. In the
current experiment we measure the interaction of the single atom with one
photon of the entangled pairs by observing quantum jumps, very similar to
Dehmelt's proposal for spectroscopy on highly forbidden transitions
\cite{Dehmelt1975BAPSv20p60}. In contrast to other recent work where weak
light fields interact with single atomic absorbers \cite{Tey2008NPv4p924,
Wrigge2008NPv4p60, Vamivakas2007NLv7p2892}, our SPDC photons bear the
potential of transferring their non-classical properties to the atoms; in
the framework of quantum technologies, our results form a step towards
implementing photon-to-atom entanglement transfer in quantum networking
scenarios \cite{Lloyd2001PRLv87p167903}.

\begin{center}
\begin{figure}[t]
\includegraphics[width=0.99\columnwidth]{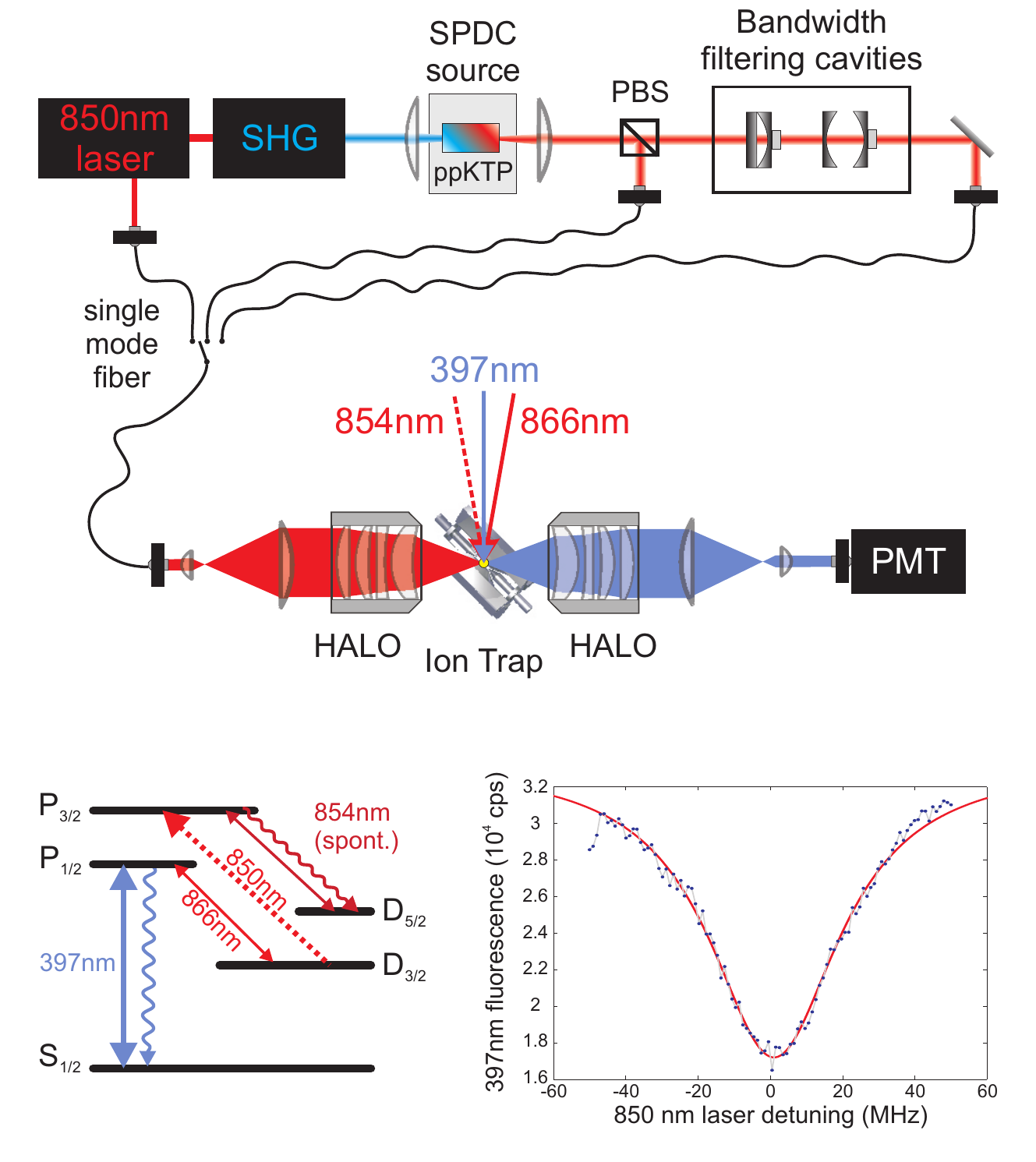}
\caption{Scheme of the experiment. The lasers at 397~nm, 866~nm, and
854~nm enter the trap from one side while fluorescence is collected
through one of the HALO lenses. The SPDC photons enter through the other
HALO lens. Either the unfiltered or the filtered arm of the source, or the
850~nm master laser, is used for spectroscopy of the ion. The relevant
levels of Ca$^+$ are shown in the bottom left panel. The bottom-right
inset shows a spectrum of 397~nm fluorescence when the 850~nm transition
is excited by the master laser of the SPDC source. We use this spectrum to
determine the 850~nm line center. A magnetic field along the optical axis
of the HALOs removes the Zeeman degeneracies and provides a quantization
axis. } \label{fig:setup}
\end{figure}
\end{center}

The experimental set-up is displayed schematically in
Fig.~\ref{fig:setup}. The ion is confined by a standard linear ion trap
surrounded by two diffraction-limited, high numerical aperture (HALO)
lenses for efficient optical access \cite{Gerber2009NJoPv11p13}.
Continuous laser excitation on the $4{\textrm S}_{1/2} - 4{\textrm
P}_{1/2}$ and $3{\textrm D}_{3/2} - 4{\textrm P}_{1/2}$ transitions at
397~nm and 866~nm, respectively, provides cooling to the Lamb-Dicke regime
and generates fluorescence which is monitored with a photon counting
detector. The type-II SPDC source produces polarisation-entangled photon
pairs at 850~nm in about 150~GHz bandwidth, which are split by a
polarising beam splitter, thus providing time- and frequency-correlated
photon pairs in two output arms. In one of the arms the photons are
filtered by two actively stabilised Fabry-Perot cavities, thus producing
narrow-band tunable photons which are matched in frequency and bandwidth
to the $3{\textrm D}_{3/2} - 4{\textrm P}_{3/2}$ transition of the Ca$^+$
ion at 849.802~nm \cite{Haase2009OLv34p55, Piro2009JPBAMOPv42p114002}.

All lasers are transfer-stabilised via optical resonators to a
saturated-absorption signal in cesium \cite{RohdeTbsvp}, including the
master laser of the SPDC source, which is tuned to the 850~nm line and
which in turn is the reference for the narrow-band filters in the filtered
arm. The inset of Fig.~\ref{fig:setup} shows a reference spectrum taken by
scanning this master laser across the 850~nm resonance while recording the
ion's fluorescence at 397~nm. The laser power is kept low such that any
light shift of the D$_{3/2}$ level is avoided. The fluorescence exhibits a
dip around the 850~nm resonance frequency, which is explained by resonant
optical pumping from D$_{3/2}$ into D$_{5/2}$ via the P$_{3/2}$ level.
Since the D$_{5/2}$ level is metastable with a lifetime of $\sim 1$~s, we
use an additional weak laser, tuned near the $3{\textrm D}_{5/2} -
4{\textrm P}_{3/2}$ transition at 854~nm, to pump the ion back into the
fluorescence cycle.

Without the 854~nm laser, weak resonant excitation at 850~nm produces
quantum jumps, i.e.\ random switching of the ion's fluorescence between
the full rate determined by the 397~nm and 866~nm excitation, and the
background level (dark counts and stray light)
\cite{Nagourney1986PRLv56p2797, Sauter1986PRLv57p1696,
Bergquist1986PRLv57p1699}. Such a "quantum amplifier" scheme
\cite{Dehmelt1975BAPSv20p60} is capable of detecting individual absorption
events at extremely low rates, which has originally been proposed for
observing forbidden resonances, while in our case it is used for detecting
an extremely low photon flux resonant with a weak dipole-allowed
transition.

\begin{center}
\begin{figure}[b]
\includegraphics[width=\columnwidth]{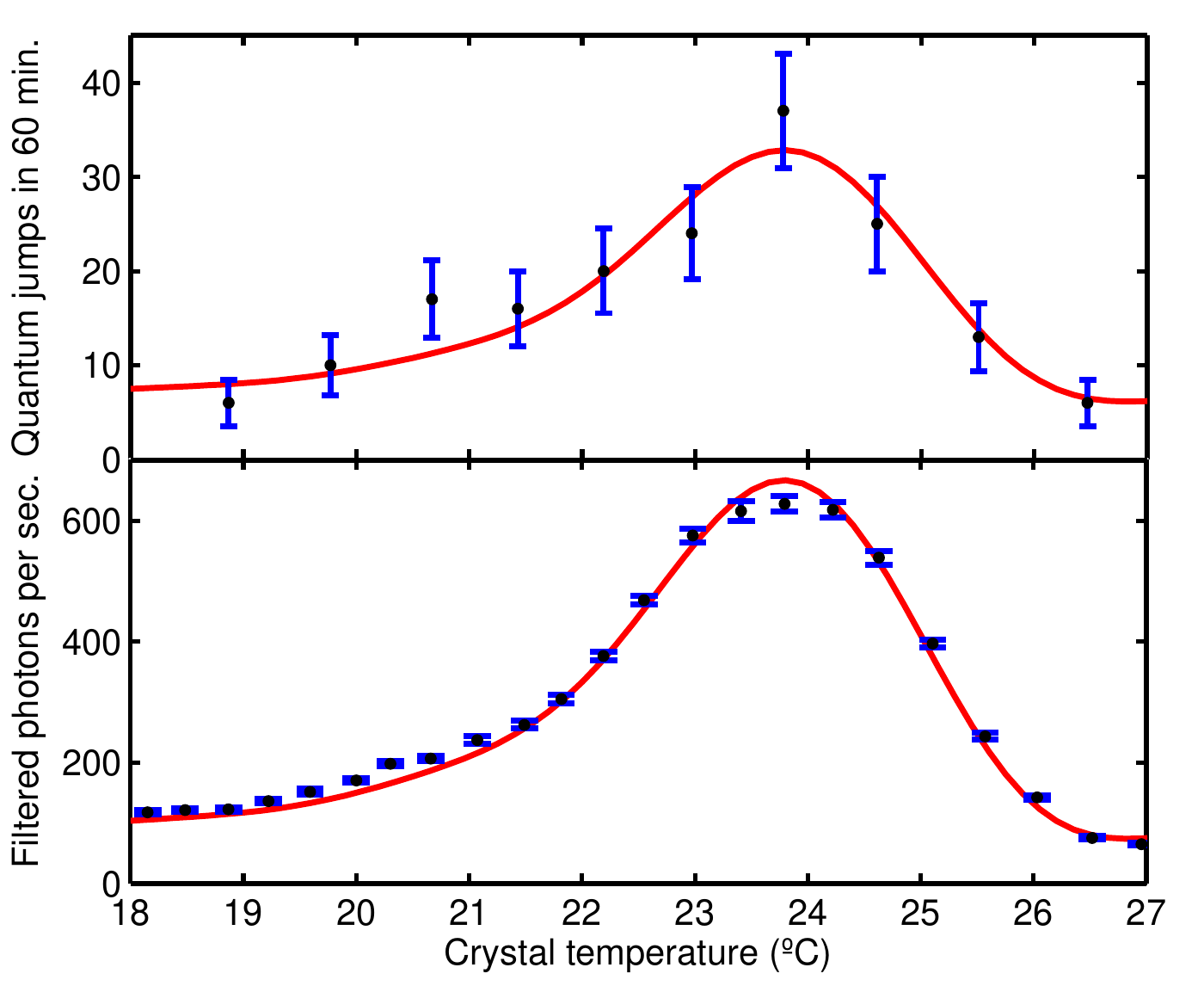}
\caption{Top: quantum jump rate induced by photons from the unfiltered
arm, as a function of the SPDC crystal temperature. Each data point
corresponds to 1~hr total measurement time, and the error bars are the
calculated Poissonian ($\sqrt{n}$) deviations. Bottom: count rate at the
output of the filtered arm, for comparison. Here the filters act as an
artificial ion with the same reson one. Points are averages of 120
measurements of 1~s each; error bars indicate the standard deviation of
the mean. In both plots the line is a theoretical calculation
\protect{\cite{Molina}}. } \label{fig:unfiltered}
\end{figure}
\end{center}

In the first quantum jump measurement, the photons from the unfiltered
output arm of the SPDC source are sent to the ion. Like the 850~nm laser
they induce optical pumping into D$_{5/2}$, but now, with the 854~nm laser
switched off, the ion remains in the metastable level for an average time
of 1.2~s before returning to the ground state by spontaneous emission. We
find a rate of about 0.7 on-off jumps per minute induced by the unfiltered
SPDC photons, the background rate without the SPDC light being 0.09/min.
Since the unfiltered arm provides photons in about 200~GHz bandwidth,
there is no narrowband spectral dependence of the quantum jump rate.
Instead, we measured the spectral variation with the temperature-dependent
emission spectrum of the SPDC source. The emission peak shifts by about
-59~GHz/$^\circ$C, such that by changing the temperature of the
down-conversion crystal over $\sim 10^\circ$C we scan the whole SPDC
bandwidth over the resonance. The result is displayed in
Fig.~\ref{fig:unfiltered}. The observed spectral (i.e.\ temperature)
dependence of the quantum jump rate, Fig.~\ref{fig:unfiltered} (top),
agrees well with the measured rate of photons within the absorption
bandwidth, Fig.~\ref{fig:unfiltered} (bottom), which is recorded by using
the tuned filters in the other arm to simulate the ion's narrow-band
absorption window.

The maximum observed jump rate is also close to what we expect, estimated
as follows. From the characterization of the SPDC source the spectral flux
of photons in the unfiltered arm is known to be around 250/(s~MHz)
\cite{Piro2009JPBAMOPv42p114002} such that within the Lorentzian
absorption line of 22~MHz bandwidth about 10$^4$ photons/s impinge on the
ion. Reducing factors are (i) the population of about 0.6 of the D$_{3/2}$
level, set by the excitation conditions on 397~nm and 866~nm; (ii) the
absorption strength on the ${\textrm D}_{3/2} - {\textrm P}_{3/2}$
transition, which contributes only $\sim 0.7\%$ to the total dipole
coupling strength of the P$_{3/2}$ level; (iii) the branching ratio for
decay of P$_{3/2}$ into D$_{5/2}$, about 5.9\%; (iv) the polarization
matching between the SPDC photons and the transition, which contributes a
reduction by 1/3; and (v) the geometric overlap of the incoming light with
the radiation pattern of the ${\textrm D}_{3/2} - {\textrm P}_{3/2}$
dipole \cite{Sondermann2007APBv89p489}, about 2\% \footnote{This number is
estimated from the efficiency with which we obtain the inverse process of
coupling the single-ion fluorescence into a single-mode optical fiber.}.
From these numbers, about 1~jump in 100~s is predicted.

\begin{center}
\begin{figure}[t]
\includegraphics[width=\columnwidth]{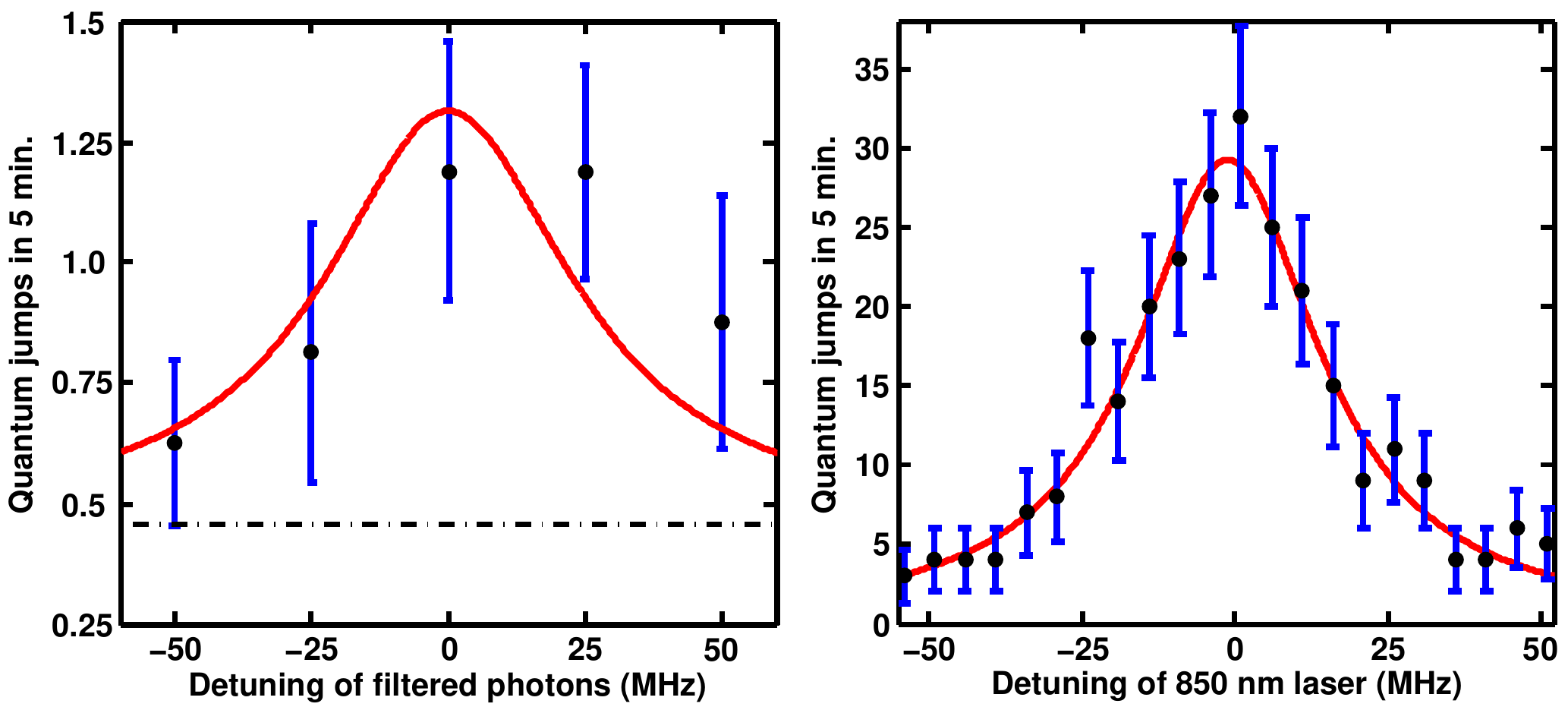}
\caption{Left: quantum jump rate induced by photons from the filtered arm,
as a function of the filter frequency. Points are averages of 16
measurements of 5~min each, error bars indicate the standard deviation of
the mean. Right: for comparison, quantum jump rate induced by the strongly
attenuated master laser, based on 5~min of data acquisition; error bars
are the calculated Poissonian deviations. The curve in the right-hand
display is a Lorentzian fit; the curve in the left-hand display is the
convolution of this Lorentzian with the spectral distribution of the
filtered photons. The background (dashed line) is determined
independently. } \label{fig:filtered}
\end{figure}
\end{center}

In the second quantum jump measurement, we make the ion interact with the
narrow-band output of the filtered arm, with a measured bandwidth of
22~MHz and stabilised to the frequency of the master laser (see
Fig.~\ref{fig:setup}). The narrow-band photons are tuned across the 850~nm
resonance by shifting with an acousto-optical modulator the master laser
frequency, to which the filter cavities are referenced. The results are
displayed in Fig.~\ref{fig:filtered}. The resonance is clearly visible,
and the linewidth corresponds to what is expected from convoluting the
22~MHz bandwidth of the SPDC photons with the spectroscopic resonance
width of 36~MHz, measured with a narrow-band laser (also shown in
Fig.~\ref{fig:filtered}) \footnote{The linewidth of 36~MHz results from
the natural width of the P$_{3/2}$ level, 25~MHz, and the Zeeman splitting
of the P$_{3/2}$ and D$_{3/2}$ manifolds, created by a magnetic field of
about 3~G.}. The maximum rate is lower than what was achieved with the
unfiltered photons, due to the attenuation of the filters ($\sim 50\%$)
and the mismatch between the bandwidth of the filtered photons and the
transition linewidth.

In summary, combining a single-ion trap equipped with high-NA optical access and a
tunable, narrow-band spontaneous parametric down-conversion source of photons, we have
observed tunable, resonant interaction between a single trapped ion and single photons
from the source. We employed a quantum jump detection scheme which allows for observing
individual interaction events at rates on the order of 1/min. In future extensions of
this work we will implement measures to increase the interaction efficiency and make use
of the correlations and the entanglement of the photon pairs to investigate
photon-to-atom entanglement transfer.

We acknowledge support from the European Commission (SCALA, contract
015714; EMALI, MRTN-CT-2006-035369), the Spanish MICINN (QOIT,
CSD2006-00019; QNLP, FIS2007-66944), and the Generalitat de Catalunya
(2005SGR00189; FI-AGAUR fellowship of C.S.). We thank G.\ Molina for the
curve in Fig.~\ref{fig:unfiltered} and for helpful remarks.



\end{document}